# Graph Learning Approaches to Recommender Systems: A Review

Shoujin Wang[1], Liang Hu[2], Yan Wang[1], Xiangnan He[3], Quan Z. Sheng[1], Mehmet Orgun[1], Longbing Cao[2], Nan Wang[1], Francesco Ricci[4], Philip S. Yu[5]

[1]Macquarie University
[2]University of Technology Sydney
[3]University of Science and Technology of China
[4]Free University of Bozen-Bolzano, [5]University of Illinons at Chicago
{shoujin.wang, yan.wang}@mq.edu.au, milkrain@gmail.com

## Abstract

Recent years have witnessed the fast development of the emerging topic of Graph Learning based Recommender Systems (GLRS). GLRS mainly employ the advanced graph learning approaches to model users' preferences and intentions as well as items' characteristics and popularity for Recommender Systems (RS). Differently from conventional RS, includeng content based filtering and collaborative filtering, GLRS are built on simple or complex graphs where various objects, e.g., users, items, and attributes, are explicitly or implicitly connected. With the rapid development of graph learning, exploring and exploiting homogeneous or heterogeneous relations in graphs is a promising direction for building advanced RS. In this paper, we provide a systematic review of GLRS, on how they obtain the knowledge from graphs to improve the accuracy, reliability and explainability for recommendations. First, we characterize and formalize GLRS, and then summarize and categorize the key challenges in this new research area. Then, we survey the most recent and important developments in the area. Finally, we share some new research directions in this vibrant area.

## 1 Introduction

Recommender Systems (RS), as one of the most popular and important applications of Artificial Intelligence (AI), have been widely adopted to help the users of many popular web content sharing and e-Commerce to more easily find relevant content, products or services. Meanwhile, Graph Learning (GL), i.e., machine learning on graph structure data, as an emerging technique of AI, has been developing rapidly and has shown great promise in recent years. Benefiting from the capability of GL to learn relational data, an emerging RS paradigm built on GL, namely Graph Learning based Recommender Systems (GLRS), has been proposed and studied extensively in the last few years. This motivates us to systematically review the challenges and progress in this area to provide a comprehensive picture of GLRS.

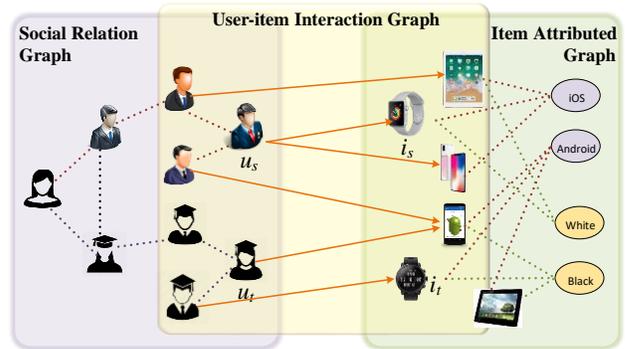

Figure 1: Some typical graphs in RS

*Motivation: why graph learning for RS?*
Most of the data in RS has essentially a graph structure. In the real world, most of the objects around us are explicitly or implicitly connected with each other; in other words, we are living in a world of graphs. Such characteristic is even more obvious in RS where the objects here considered including users, items, attributes, etc. are tightly connected with each other and influence each other via various relations [Hu *et al.*, 2014], as shown in Figure 1. In practice, various kinds of graphs exist in the data used by RS, and significantly contribute to the recommendations. This intrinsic data characteristic drives the necessity to consider the complex inter-object relations when making recommendations.

Graph learning has the strength in learning complex relations. As one of the most promising machine learning techniques, GL has shown great potential in obtaining knowledge embedded in different kinds of graphs. To be specific, many GL techniques, such as random walk and graph neural networks, have been developed to learn the particular type of relations modeled on graphs, and have demonstrated to be quite effective. Consequently, employing GL to model various relations in RS is a natural and wise choice.

Graph learning helps in building explainable RS. Nowadays, in addition to accuracy, explainability of recommendations has attracted more and more attention from both academia and industry. This trend is even more obvious in recent years since RS are dominated by deep learning techniques which are often run under black-box working mechanisms. Benefiting from the causal inference capability of GL for relations, GLRS can easily support the explanation of the recommendation results in terms of the inferred relationships between different objects involved in RS. This actually greatly promotes the development of the explainable RS.

*Formalization: how does graph learning can help RS?*
To date, there is no unified formalization of all GLRS. In fact, there are different and specific implementations applying different models on different data with specific characteristics. Consequently, we generally formalize GLRS from a highlevel perspective.

Given a dataset, consider the graph G = {V,E} where the objects, e.g., users, items, are considered as nodes in set V and relations, e.g., purchase history, social relations, between them are edge in set E. Then, GLRS use the graph G as input to generate the corresponding recommendation result R, by modelling the topological and content information on G. Formally,

$$R = \mathrm{argmax} f(G) \qquad (1)$$

Depending on the specific data and recommendation scenarios, the graph G may be homogeneous or heterogeneous, static or dynamic, while R can be in various forms, e.g., predicted ratings or ranking over items. The optimization objectives are also different in specific GLRS: they could be the maximal choice utility according to graph topological structure or the maximal probability to form links between nodes.

*Contributions.* The main contributions of this work are summarized below:

- We systematically analyze the key challenges commonly existing on various graphs in GLRS and categorize them from a data driven perspective, providing a new view to deeply understand the characteristics of GLRS.

- We summarize the current research progress in GLRS by systematically categorizing the state-of-the-art works from a technical perspective.

- We share and discuss some open research directions of GLRS as references for the community.

## 2 Data Characteristics and Challenges

In general, there are many different objects that an RS can consider, e.g., users, items, attributes, contexts, etc., and nearly all of them are connected with each other with certain types of links [Hu *et al.*, 2019], e.g., social relations over users, interactions between users and items. These various relations essentially result in natural graphs in RS. Moreover, the diversities of objects and the relations between them lead to different types of graphs which correspond to different challenges. In this section, we will start from a data-driven perspective to systematically analyze the data complexities and characteristics in RS, and accordingly demonstrate the challenges when building RS on different graphs with different specific characteristics. A brief summary is provided in Table 1.

### 2.1 RS on Tree Graphs

Usually, all the items in a transaction dataset for recommendations are organized with a hierarchy which is also called a tree graph, according to a certain attribute of items, e.g., the category [Huang *et al.*, 2019]. For instance, the items sold on Amazon.com are first categorized into different categories (e.g., electronic products and sport products), and then each category is divided into several sub-categories (e.g., wireless products and computers in electronic products) while each sub-category includes multiple items (e.g., iPhone XR, Huawei Watch GT in wireless products). Such hierarchical graph essentially reveals the rich relations behind items [Gao *et al.*, 2019], for example, two items (e.g., an iPhone XR and an accessory) from different but closely related categories are very likely to have complementary relations w.r.t. their functions. Such hierarchy relations between items can greatly improve the performance of recommendations, e.g., to avoid repeatedly recommendding similar items from the same subcategory to a user and thus to diversify the recommendation list. Consequently, *how to effectively learn such hierarchy relations between items and incorporate them into the subsequent recommendation task* is an important challenge.

### 2.2 RS on Unipartite Graphs

In RS, at least two homogeneous unipartite graphs may be defined, one for users and the other for items. Specifically, on the one hand, the online or offline social relations between users constitute a homogeneous social graph for users [Bagci and Karagoz, 2016]; on the other hand, the co-occurrence of items in the same shopping basket or session connects all the items in a transaction data together and thus results in a homogeneous session graph for items [Xu *et al.*, 2019]. Users in a social graph usually influence each other w.r.t. their preference on items as well their shopping behaviours. Therefore, it is necessary to take the social influence into account when making recommendations. Moreover, such social influence usually propagates on the social graph and thus should have a cascading impact on the recommendation results. Subsequently, *how to learn the inter-user social influence and its propagation over the user graph for recommend-ations* becomes a specific challenge. Similarly, the co-occurrence relations between items usually not only reflect certain latent relations between items, e.g., the complementary or competitive relations w.r.t. the items' functions, but also reveal some shopping patterns of users. Hence, incorporating the co-occurrence relations between items over the session graph helps generate more accurate recommendations. This brings another specific challenges: *how to substantially capture the inter-item relations on the item graph and appropriately utilize them to improve the accuracy of recommendations.*

### 2.3 RS on Bipartite Graphs

The interactions (e.g., clicks, purchases) connecting users and items are the core information for an RS and all of them together naturally form a user-item bipartite graph. According to the number of types of interactions considered on the graph, the interactions modeled in the bipartite graph can be homogeneous (only one type of interactions) or heterogeneous (multi-types of interactions). In general, the task to recommend items to a certain user can be seen as the link prediction over a user-item bipartite graph, namely, given the known edges in the graph to predict the possible unknown ones [Li and Chen, 2013]. In such a case, a typical challenge is *how to learn the complex user-item interactions on a graph with homogeneous interactions as well as the comprehensive relations between these interactions for recommendations.* Moreover, *how to capture the influence between different types of interactions on a graph with heterogeneous interactions*, e.g., the influence of clicks on purchases, to provide richer information for generating more accurate recommendations is another bigger challenge.

### 2.4 RS on Attributed Graphs

In addition to the aforementioned homogeneous user/item graph, heterogeneous user graph, item graph or user-item interaction graph is also very common in RS. For instance, in a heterogeneous user graph, there are at least two different types of edges representing different relations respectively: one indicates the social relations between users while the other indicates that a user has a certain attribute value (e.g., male), and the users sharing the same attribute value are indirectly connected on the graph [Yin et al., 2010]. Both the users' attribute values and the indirect connections built on them are of great significance for improving the performance of both friend recommendations [Verma et al., 2019] and social recommendations (recommending items while incorporating social relations) [Fan et al., 2019] by providing additional information to better capture personalized user preference and the inter-user influence.

| GLRS cases | Graph instances | Recommendation tasks | Approaches |
| --- | --- | --- | --- |
| RS on tree graphs | Item hierarchical graphs | Rating prediction | Knowledge graph [Gao et al., 2019] |
| RS on unipartite graphs | Social graphs of users, Session graphs of items | Friend recommendations, Next-item recommendations | Random walk [Bagci and Karagoz, 2016], Graph neural networks [Wu et al., 2019b; Xu et al., 2019] |
| RS on bipartite graphs | User-item interaction graphs based on ratings, purchase, ectc. | Top-n item recommendations | Random walk [Li and Chen, 2013] |
| RS on attributed graphs | User attributed graphs, Item attributed graphs | Friend recommendations, Social recommendations | Graph representation learning [Verma et al., 2019], Graph neural networks [Fan et al., 2019] |
| RS on complex heterogeneous graphs | User-item interaction graphs combined with social relations or item features | Social recommendations, Rating prediction, Top-n item recommendations | Knowledge graph [Palumbo et al., 2017], Graph neural networks [Han et al., 2018] |
| RS on multi-source heterogeneous graphs | Attributed multiplex heterogeneous graphs | Rating prediction, Top-n item recommendations | Graph representation learning [Cen et al., 2019] |

Table 1: A summary of GLRS cases from the data characteristics perspective

This actually brings the challenges of *how to model the different kinds of relations as well as the mutual-influence between them on a heterogeneous user graph*, and then *how to appropriately integrate them into recommendation tasks* [Wang et al., 2019a, Wang et al., 2019b].

Similarly, the co-occurrence relations between items and the item-attribute values relations form a heterogeneous item graph. Both relations are important to understand the distributions, occurrence patterns, and the intrinsic nature of items and thus benefit the recommendations. Therefore, *how to effectively model such heterogeneous relations on a heterogeneous item graph to improve recommendation performance* becomes another challenge in this branch.

### 2.5 RS on Complex Heterogeneous Graphs

To address the sparsity issue in user-item interaction data for better understanding of user preference and item characteristics, auxiliary information like social relations, or item features are often combined with user-item interaction information for better recommendations. On the one hand, to consider the inter-user influence w.r.t.

preference on items, social relations between users are usually combined together with user-item interactions, to build the so-called social RS [Guy, 2015]; on the other hand, to characterize items deeply, item features are often combined together with user-item interactions to provide recommendations on cold-start items [Palumbo *et al.*, 2017; Han *et al.*, 2018]. Such combination of two types of heterogeneous information for recommendations results in two heterogeneous graphs: one is the bipartite graph based on user-item interactions, and the other is the social graph between users or the item-feature graph. The shared users or items in both graphs serve as the bridge to connect them. Social relations or item features are quite important to deeply understand users by considering the preference propagation over them, or to better characterize items by considering their natural attributes. However, it is quite challenging to *enable the heterogeneous information from the two graphs to communicate with each other appropriately and to be combined inherently to benefit the recommendation task.*

## 2.6 RS on Multi-source Heterogeneous Graphs

To effectively address the ubiquitous data sparsity and coldstart issues to build more robust and reliable RS, except for the user-item interactions, a lot of relevant information that has great explicit or implicit impact on recommendations from multi-sources can be effectively leveraged and integrated into RS [Cen *et al.*, 2019]. For instance, the user profiles from user information table, social relations from the online or offline social networks, item features from item information table, item co-occurrence relations from transaction table, etc., can be leveraged simultaneously to help with the better understandding of user preference and item characteristics for improving recommendations. Accordingly, multiple heterogeneous graphs are jointly built for recommendations: the user-item interaction based bipartite graphs providing the key information for modeling user choices, user attributed graphs and social graphs providing auxiliary information of users while item attributed graphs and item co-occurrence based graphs providing auxiliary information of items. Generally speaking, the more we know about a user's preference and shopping behaviours, together with a item's characteristics and popularity from all available information, the better recommendations we can make. However, the information on different graphs is relatively disconnected and cannot be immediately used due to this heterogeneity, thus *how to exploit together different graphs to complement each other and benefit recommendations* is the first challenge. Moreover, more heterogeneous graphs mean higher risk that there may be noise, or even contradictions between different graphs. So *how to extract coherent information and reduce the noisy and incoherent information from multi-source heterogeneous graphs to improve the downstream recommendations* is another big challenge.

## 3 Graph Learning Approaches to RS

After having discussed the different types of graphs with different data characteristics and challenges in building RS, we now examine how to address these specific challenges and to what extent they can be addressed. In this section, we will first provide a categorization of the solutions to these challenges for building GLRS from the technical perspective, and then discuss the progress achieved in each category .

The categorization of approaches to GLRS is presented in Figure 2. GLRS are first divided into four categories, and some categories (e.g., Graph neural network approach) are further divided into multiple sub-categories. In General, these categories change from simple to complex and are reported successively. Next, we summarize the research progress in each of these four categories.

### 3.1 Random Walk Approach

Random walk based RS have been extensively studied in the past 15 years and have been widely employed on various graphs (e.g., social graphs between users, co-occurrence graph between items) to capture complex relations between nodes for recommendations. Generally, random walk based RS first let a random walker walk on the constructed graph with users and/or items as its nodes with a predefined transition probability for each step to model the implicit preference or interaction propagation among users and items, and then take the probability the random walker lands on nodes after certain steps to rank these candidate nodes for recommendations. Benefiting from its particular work mechanism, random walk based RS are good at capturing the complex, higher-order and indirect relations among a variety of nodes (e.g., users and items) on the graph and thus can address important challenges in homogeneous or heterogeneous graphs to generate recommendations.

There are different variants of random walk based RS. Besides the basic random walk based RS like [Baluja *et al.*, 2008], random walk with restart based RS [Bagci and Karagoz, 2016; Jiang *et al.*, 2018] is a another representative type of random walk based RS. It sets a constant probability to jump back to the starting node in each transition and it is generally used in graphs containing many nodes to avoid moving out of the particular context of the starting node. In addition, transition probability is one of the key factors determining the recommendation results. To provide more personalized recommendations, some random walk based RS [Eksombatchai *et al.*, 2018] calculate a user-specific transition probability for each step. Other typical applications of random walk based RS including ranking items w.r.t. their importance on item-item co-viewed graph [Gori *et al.*, 2007], recommending top-n items to users by simultaneously modeling the user-item interactions on user-item bipartite graph while using item-item proximity relations to guide the transitions [Nikolakopoulos and Karypis, 2019].

Although widely applied, the drawbacks of random walk based RS are also obvious: (1) they need to generate ranking scores on all candidate items at each step for each user, and thus they are hard to be applied on large-scale graphs due to the low efficiency, (2) unlike most learning-based paradigm, random walk based RS are heuristic-based, lacking model parameters to optimize the recommendation objective, which greatly reduces the recommendation performance.

### 3.2 Graph Representation Learning Approach

Graph representation learning is another effective technique to analyze the complex relations embedded on graphs and has been developing rapidly in recent years. It maps each node into a latent low-dimension representation such that the graph structure information is encoded into it. Researchers introduced graph representation learning into RS to model the complex relations between various nodes (e.g., users, items) for the subsequent recommendations and thus built Graph Representation Learning based RS (GRLRS). According to the specific approaches used for representation learning, GRLRS can be generally divided into three classes: (1) Graph Factorization Machine based RS (GFMRS), (2) Graph Distributed Representation based RS (GDRRS), and (3) Graph Neural Embedding based RS (GNERS). Next, we present the progress achieved in each class respectively.

Graph Factorization Machine based RS (GFMRS). GFMRS first employs factorization machines (e.g., matrix factorization) to factorize the inter-node commuting matrix based on meta-path on the graph to obtain the latent representations of each node (e.g., a user or an item), which will be used as the input of the subsequent recommendation task [Wang *et al.*, 2019d]. By doing so, the complex relations between nodes embedded in the graph are encoded into the latent representations to benefit the recommendations. Due to the capability to handle the heterogeneity of nodes, GFMRS have been widely applied to capture the relations between different types of nodes, e.g., users and items. Although simple and effective, such models may easily suffer from the sparsity of the observed data and thus it is hard to achieve ideal recommendations.

Graph Distributed Representation based RS (GDRRS). Differently from GFMRS, GDRRS usually follow Skip-gram model [Mikolov *et al.*, 2013] to learn a distributed representation for each user or item in a graph to encode the self information of the user or item and its adjacent relations into a low-dimensional vector [Shi *et al.*, 2018], in preparation for the subsequent recommendations. Specifically, GDRRS usually first use random walk to generate a sequence of nodes that co-occurred in one meta-path and then employ the skipgram or similar models to generate node representations for recommendations. Taking the advantage of its powerful capability to encode the inter-node connections on graph, GDRRS are widely applied on both homogeneous or heterogeneous graphs to capture the relations between various objects in RS [Cen *et al.*, 2019]. Without deep or complex network structure, GDRRS have shown great potential in recent years due to its simplicity, efficiency and efficacy [Wang et al., 2020a, Wang et al., 2020b].

Graph Neural Embedding based RS (GNERS). GNERS generally utilize neural networks, like Multilayer-perceptron (MLP), to learn the embeddings of users or items in a graph and then use the learned embeddings for recommendations. Neural embedding models are easy to integrated together with other downstream neural recommendation models (e.g., RNN based ones) to build an end-to-end RS, which can jointly train the two models together for better optimization [Han *et al.*, 2018]. To this end, GNERS have been broadly applied on a variety of graphs like attributed graphs [Han *et al.*, 2018], heterogeneous graphs [Hu *et al.*, 2018], multi-source heterogeneous graphs [Cen *et al.*, 2019], etc., to resolve various challenges for recommendations.

### 3.3 Graph Neural Network Approach

In recent years, Graph Neural Networks (GNN) which apply neural networks on graph data, are developing rapidly and have shown great potential in addressing a variety of challenges on various graphs. Benefiting from the strength of GNN, a lot of GNN based RS are proposed by introducing GNN to address different challenges in GLRS. In particular, GNN based RS can be mainly categorized into three classes according to the specific GNN models that are utilized: (1) Graph ATtention network based RS (GATRS), (2) Gated Graph Neural Network based RS (GGNNRS), and (3) Graph Convolutional Network based RS (GCNRS).

Graph ATtention network based RS (GATRS). Graph ATtention networks (GAT) introduce attention mechanism into GNN to discriminatively learn the different relevance and influence degrees of other users or items w.r.t. the target user or item on a user or item graph. Specifically, attention weights are learned to attentively integrate the information from neighbours into the representation of the target user or item. GATRS are built on GAT for more precisely learning the inter-user or item relations for the subsequent recommendations. In such a case, the influence from those more important users or items w.r.t. a specific user or item is emphasized, which is more in line with the real-world cases and thus benefits improving the recommendations. Due to the good discrimination capability, GAT are widely used in different kinds of graphs including social graphs [Fan *et al.*, 2019], item session graphs [Xu *et al.*, 2019], and knowledge graphs [Wang *et al.*, 2019c], to build various GATRS with good recommendation performance.

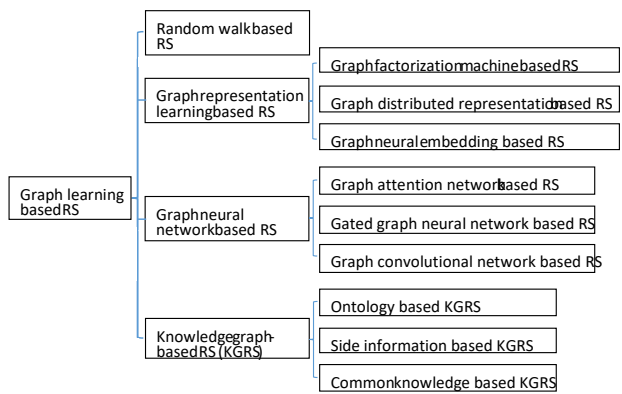

Figure 2: A categorization of GLRS approaches from the technical perspective

Gated Graph Neural Network based RS (GGNNRS).
Gated graph neural networks (GGNN) introduce the Gated Recurrent Unit (GRU) into GNN to learn the optimized node representations by iteratively absorbing the influence from other nodes in a graph to comprehensively capture the internode relations. GGNNRS are built on GGNN to learn the user or item embeddings for recommendations by comprehensively considering the complex inter-user or inter-item relations embedded on the corresponding user or item graph. Due to the strength of capturing the complex relations on a graph, GGNN are widely used to model the complex transitions between items in a session graph for session-based recommendations [Wu *et al.*, 2019b], or to model the complex interactions between different categories of fashion products for fashion recommendations [Cui *et al.*, 2019, Wang et al., 2019c], and have achieved superior recommendation performance.

Graph Convolutional Network based RS (GCNRS). Graph Convolutional Networks (GCN) generally learn how to iteratively aggregate feature information from local graph neighborhoods using neural networks by leveraging both graph structure and node feature information. In general, by utilizing the convolution and pooling operations, GCN is capable of learning informative embeddings of users and items by effectively aggregating information from their neighborhoods in graphs. GCNRS are built on GCN to learn the uer or item embeddings in a graph while comprehensively exploiting the complex relations between users or/and items as well as their own information at the same time for recommendations [Ying *et al.*, 2018]. Thanks to the powerful feature extraction and learning capability, particularly the strength in combining the graph structure and node content information, GCN are widely applied to a variety of graphs in RS to build GCNRS and are demonstrated to be of great promising. For instance, GCN are used for influence diffusion on social graphs in social recommendations [Wu *et al.*, 2019a], mining the hidden user-item connection information on user-item interaction graphs to alleviate the data sparsity issue in collaborative filtering [Wang *et al.*, 2019a], and capturing inter-item relatedness by mining their associated attributes on knowledge graph built on item attributes [Wang *et al.*, 2019b].

### 3.4 Knowledge-Graph Approach

Knowledge-Graph based RS (KGRS) generally build a Knowledge-graph (KG) based on external knowledge, e.g., side information, to explore implicit or high-order connectivity relations between users or items to enrich their representations and thus improve recommendation performance. More significantly, due to the utilization of additional knowledge, KGRS are capable of providing better understanding of user behaviours and item characteristics, which leads to more explainable recommendations [Wang *et al.*, 2018a]. KGRS mainly focus on the construction of the KG in the early stage of an RS, while various existing techniques including factorization machines, graph neural networks, are employed to extract information from the constructed KG and integrated it into the subsequent recommendations. According to the knowledge used to build the KG, KGRS can be generally divided into three classes, which will be presented successively below.

Ontology based KGRS (OKGRS). OKGRS build a hierarchical KG based on the ontology of users or items to represent the hierarchical belonging relations in tree-like graphs. A typical example of hierarchical KG is the tree graph used in Amazon.com, where the category of products is utilized to organize all the items on sale on the platform. In this graph, the root nodes indicate the most coarse-grained category like food while the leaf nodes represent the specific items like bread. In recent years, OKGRS have been widely studied to enhance the explainability of recommendations, e.g., using it to extract multi-level user interest from the item ontology graph [Gao *et al.*, 2019, Wang et al., 2017].

Side information based KGRS (SKGRS). SKGRS build a KG based on the side information of users or items (e.g., attributes of items) to discover the implicit connections between them for improving recommendation performance. For example, the implicit connections between the items which share the same attribute values (e.g., drink) provide additional information to understand the inter-item relations in RS. As a result, KG are widely used in collaborative filtering to enrich the item representations by incorporating additional information for improving the recommendations [Wang *et al.*, 2019c].

Common knowledge based KGRS (CKGRS). CKGRS mainly build the KG based on common knowledge, e.g., the general semantic extracted from online texts, domain knowledge, etc. CKGRS incorporate the external implicit relations between recommended products or services that are extracted from common knowledge for improving recommendations. Therefore, they are widely applied in news recommendations to discover latent knowledge-level connections among news [Wang *et al.*, 2018b] and e-commerce to infer users' latent needs [Luo *et al.*, 2019].

## 4  Open Research Directions

GLRS are developing rapidly. Although great progress has been achieved as illustrated in the last section, some challenges still remain not well resolved and thus require more efforts. By linking the research progress already achieved and the demonstrated challenges in this area, we have identified further open research directions to be discussed below.

Dynamic-graph learning for RS. In real-world RS, the objects including users and items as well as the relations between them are changing over time, which results in dynamic graphs instead of static ones. Such dynamics could have a significant impact on the recommendation results or even change the recommendations over time. However, such a case is often ignored or less studied in the existing GLRS. Therefore, it is an important direction for future work to learn on dynamic graphs for RS.

Casual inference based graph learning for RS. Casual inference is a main technique to discover the casual relations between objects or actions. Although some progress has been achieved in explainable RS, we are still far away from substantially and completely understanding the reasons and intents behind users' choices, which is very critical for making reliable and explainable recommendations. To this end, it is another promising direction to incorporate casual inference into GLRS to build more advanced and explainable RS. Multi-source and multi-modality graph learning for RS. In reality, the data points for recommendations could be from multiple sources with various modalities, but they are interrelated and collaboratively contribute to the recommendations. In addition, some types of data points may bring some noise to others and there may be even some contradictions among these diverse data points. Consequently, it deserves further explorations how to build an effective and efficient GLRS on multi-source and multi-modality graphs.

Large scale and real-time graph learning for RS. A typical issue in real world is that the datasets used for RS are so large, leading to high cost in terms of both time and space for RS. Such issue is even more evident in GLRS since the graph structure data is usually larger and requires more time and space to process, let alone perform complex machine learning techniques on it to generate recommendations. To this end, it is necessary to further study more advanced GLRS which enable large-scale and real-time computations to generate recommendations.

## 5  Conclusions

As one of the most important and practical applications of Artificial Intelligence (AI), Recommender Systems (RS) can be found nearly each corner of our daily lives. Graph Learning (GL), as one of the most promising AI techniques, have show great strength in learning the complex relations among various objects involved in an RS. This gives birth to a totally new RS paradigm: Graph Learning based Recommender Systems
(GLRS), which is of great potential to be the next-generation RS. It is our hope that this review provides an overview of the challenges and the recent progress as well as some future directions in GLRS to both the academia and industry.